\documentclass[ 12pt,draftclsnofoot,onecolumn]{IEEEtran}

\usepackage[utf8]{inputenc}

\usepackage{float}

\usepackage{rotating}

\usepackage{multirow} 

\usepackage{bigstrut}

\usepackage{graphicx}

\usepackage{xcolor}

\usepackage{cite}
\usepackage{authblk}

\graphicspath{ {./imgs/} }
\DeclareGraphicsExtensions{.pdf,.png}
\providecommand{\keywords}[1]
{
  \small    
  \textbf{\textit{Keywords---}} #1
}

\title{Backflash Light as a Security Vulnerability in Quantum Key Distribution Systems}
\author{Ivan Vybornyi, Abderrahmen Trichili, and Mohamed-Slim Alouini*}
\affil{\textit{Computer, Electrical and Mathematical Sciences $\&$ Engineering}\\ \textit{King Abdullah University of Science and Technology, Thuwal, Makkah Province, Kingdom of Saudi Arabia.}\\}

\begin{document}
\maketitle
\begin{abstract}
Based on the fundamental rules of quantum mechanics, two communicating parties can generate and share a secret random key that can be used to encrypt and decrypt messages sent over an insecure channel. This process is known as quantum key distribution (QKD). Contrary to classical encryption schemes, the security of a QKD system does not depend on the computational complexity of specific mathematical problems.
However, QKD systems can be subject to different kinds of attacks, exploiting engineering and technical imperfections of the components forming the systems. Here, we review the security vulnerabilities of QKD. We mainly focus on a particular effect known as backflash light, which can be a source of eavesdropping attacks. We equally highlight the methods for quantifying backflash emission and the different ways to mitigate this effect.
\end{abstract}
\bibliographystyle{greee}
\section{Introduction}
The history of cryptography spans thousands of years, starting with Caesar cipher, which was used by the emperor to protect secret military messages by a simple encoding scheme. Since then, many other different encryption techniques were suggested to transfer information securely. However, most of these techniques were broken or proven to be critically vulnerable. Even the protocols that are widely used in our daily communication and data transfer operations, such as the RSA (Rivest–Shamir–Adleman cryptosystem) and the elliptic-curve-based protocols, are in great danger since they rely on the complexity of solving difficult mathematical problems. RSA, for example, exploits the complexity of the factorization of large integers, and elliptic-curve algorithms are based on finding discrete logarithms. In practice, complexity does not allow a classical computer to break the protocol. However, a proper large-scale quantum computer, once is built, would be able to solve the required problems much faster than a classical one and thus would allow us to crack these protocols \cite{Shor1997}. Moreover, the fact that there will be no classical algorithms for these problems created one day remains unproven \cite{Gisin2002}. In this regard, quantum-resistant methods of cryptography need to be created. One possible solution is to use complicated quantum-resistant algorithms instead of the existing ones. This solution is also known as the post-quantum cryptography and is currently being developed but seems to be far from perfect \cite{Cheng2017,Chen2016}. Another promising alternative is the quantum cryptography in which security is based on the fundamental physical laws rather than the computational complexity.
\section{Quantum Cryptography and Quantum Key Distribution}

The basic principles of quantum mechanics were established at the beginning of the XX century as a consequence of many experimental facts and shook the perceptions of everyone of physics. For instance, the fact that any measurement perturbs the system or that certain physical pairs of variables could not be in principle measured simultaneously with arbitrarily high precision at first seemed to be unnatural or even ``counter-intuitive". However, while being restricted, all these facts have a positive side for cryptography.

\begin{figure}[H]
%\label{exp_description}
\center{\includegraphics[scale=0.6]{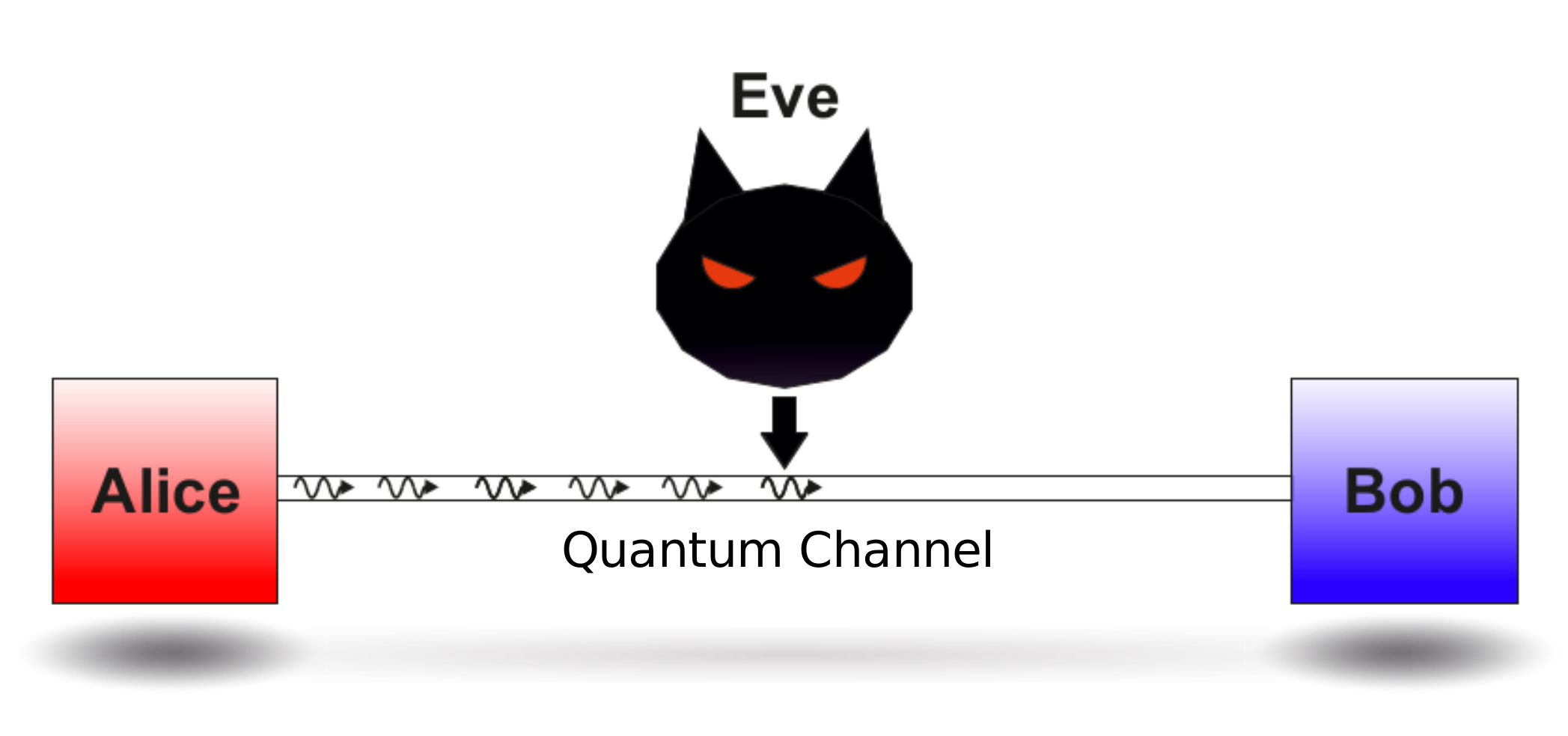}}
\caption{A typical scenario in quantum cryptography: Alice communicates with Bob and Eve attempts to eavesdrop.}
\label{Schematic}
\end{figure}

A typical cryptography scenario includes two distanced parties, commonly known as Alice and Bob, that want to transfer some secret information via a probably insecure channel. A malicious eavesdropper, referred to as Eve, intends to get the hidden information and remains unrevealed, as illustrated in Fig.~\ref{Schematic}.

The sensibility of a quantum mechanical system to any measurement can be exploited by the parties to reveal the presence of an eavesdropping Eve in the channel. The idea relies on the fact that Eve cannot get any information on what is being communicated between Alice and Bob without perturbing the state of quantum bits (qubits) transmitted through the channel. Perturbations created by Eve lead to some consequences in the communication, usually in the form of transmission errors that can be spotted by Alice and Bob.

Another interesting fact that quantum mechanics provides us with and that could be useful for cryptography is the so-called no-cloning theorem. The name says everything; in quantum mechanics, one cannot create a copy of a qubit in an arbitrary quantum state. The fact that Eve could perform bit-copying underlays many classical eavesdropping schemes; however, in the quantum-world cryptography, such schemes are irrelevant. These features, along with Heisenberg's uncertainty principle, have opened up new vistas for the secure information transfer that relies not on computational complexity but on the fundamental laws of physics. As a result,  the foundations of the new cross-discipline between quantum physics and information theory, known today as quantum cryptography, were proposed at the end of the XX century, \cite{Bennett2014}.

To date, quantum physics is not harnessed to transfer securely meaningful information itself. It is more convenient to use the quantum channel for sharing the secret key, which is subsequently used in conjunction with the traditional cryptography protocols for message encryption. The reason is that quantum effects allow Alice and Bob to reveal an eavesdropper easily only after the transfer of information. In order to detect the presence of Eve before revealing the information, quantum cryptography typically carries out the tasks of key distribution. Existing protocols of quantum information transfer also provide us only with low data rates. In this regard, quantum-based cryptography carries out the function of a key distribution mean and is referred typically to as quantum key distribution (QKD).

\section{QKD Protocols and vulnerabilities}
Today there exist dozens of different QKD protocols. They differ in the concrete physical realizations of the information channel, the detection scheme of the states of qubits, etc. The first proposed QKD protocol is BB84. BB84 was developed by Charles H. Bennet and Gilles Brassard in 1984 \cite{Bennett2014}. Within this protocol, the quantum effects allow Alice to share a random secret key with Bob to be used in conjunction with any convenient symmetric-key algorithm. Only one secret key is used in symmetrical algorithms both to encrypt and decrypt messages. Thus the task of BB84 is to transfer a truly random sequence of bits, which is then used as a key if no eavesdropping is confirmed. 

Most of the existing QKD protocols, including the BB84, implement qubits via single photons with certain polarization states. Single photons are an excellent choice since they weakly interact with the environment, and much progress has been made during the last decades in single-photon electronics. In quantum mechanics, the polarization state of a photon is described by a specific vector of a unit norm in a 2-dimensional Hilbert space. Every polarization measurement instrument has an eigenbasis of 2 orthogonal states in this space, and any measurement of the polarization state of a photon via this instrument corresponds to a projection of the polarization state vector on one of these eigenstates. Such projections manifest probabilistic behavior. In fact, the result of measurement goes with the probability determined by the components of the initial state vector of the photon in the eigenbasis of the measurement.  As an example, we consider a polarizing beamsplitter (PBS), which splits photons with random polarization into vertically and horizontally polarized ones. The eigenbasis of the PBS consists of horizontal (H) and vertical (V) polarization states. A photon with a  polarization state in H-V basis, and represented by a vector $(1,0)$ will always be registered as a photon with horizontal polarization. However, if the state of a photon in an H-V basis is $(\frac{\sqrt{2}}{2},\frac{\sqrt{2}}{2})$ then a measurement result will be entirely undetermined since the projections of state vector on the eigenstates of the measurement instrument are of equal length. So, if a large number of such measurements with the same PBS is performed under the same conditions, one will observe about half of the photons registered with a horizontal polarization state and about the second half of photons with a vertical polarization state.

\begin{figure}[H]
%\label{exp_description}
\center{\includegraphics[scale=0.6]{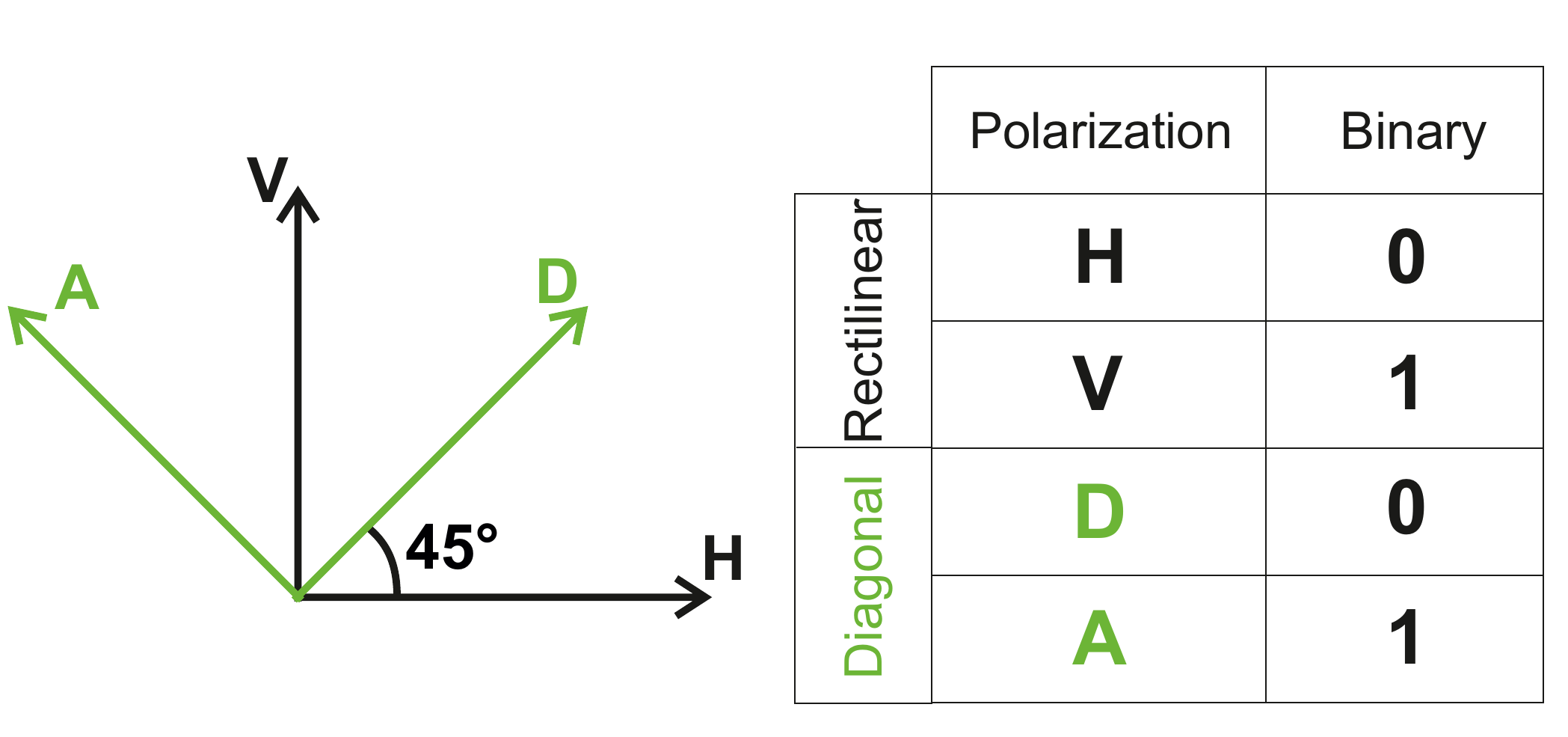}}
\caption{Diagonal (D-A) and rectilinear (H-V) bases.}
\label{Basis}
\end{figure}

\begin{figure}[H]
%\label{exp_description}
\center{\includegraphics[scale=0.23]{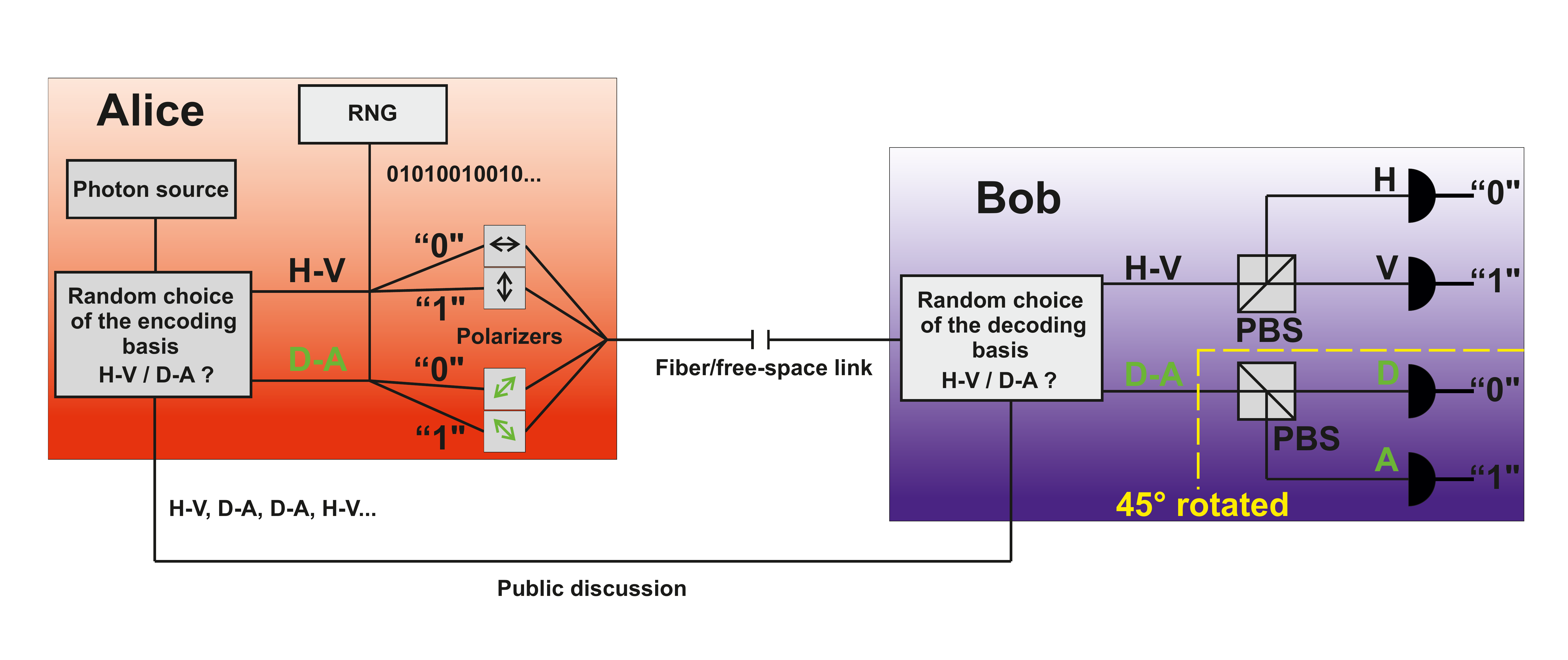}}
\caption{Schematic illustration of the BB84 protocol.}
\label{BB84Basis}
\end{figure}

Consider now a PBS with another eigenbasis and let it consist of diagonal (D) and anti-diagonal (A) polarization states. The H-V  and D-A bases are said to be `conjugate' since each vector of one basis has projections of equal length onto all vectors of the other basis. In such a situation, a photon prepared in a specific state of one basis produces entirely random measurement results when measured using an instrument with the eigenbasis formed by the vectors of another basis \cite{Bennett2014}. Suppose the setup of Alice allows her to choose a random basis (H-V or D-A) for every photon she wants to transmit. She encodes her truly random bit sequence into the polarization states of photons according to the chosen basis; 0 for H and 1 for V in the H-V basis or 0 for D and 1 for A in the D-A basis, as can be seen in Fig.~\ref{Basis}. Alice then sends the train of photons to Bob. At the side of Bob, there is a so-called passive basis choice scheme. For each incoming photon, Bob chooses randomly and independently of Alice a basis (H-V or D-A) for the measurement of polarization. The process can be done by directly sending the incoming train of photons to a 50:50 beamsplitter, which splits the photons and forwards them to the H-V or D-A measurement devices, as illustrated in Fig.~\ref{BB84Basis}. Alice and Bob then discuss via a public information channel whether the bases for detection were chosen correctly or not. The two parties agree on the qubit and get the same bit value whenever the detection has proceeded on the proper basis. They also have to disregard the qubit if the basis is wrongly chosen since H-V and D-A bases are conjugate, and nothing could be said of the original state of the photon sent by Alice after a wrong detection. In this way, two parties obtain the so-called `sifted key', which appears to be two times shorter than the original sequence of bits sent by Alice (see Table \ref{tab1}). The reason is that Bob manages to guess the basis correctly for 50\% of the transmitting photons, assuming that there is no eavesdropping, and the channel is perfect.

\begin{table}[htbp]
  \centering
  \caption{The BB84 Key Obtaining Procedure.}
    \begin{tabular}{|r|l|l|l|l|l|l|l|l|l|}
    \hline
    \multicolumn{1}{|c|}{\multirow{3}[6]{*}{\begin{sideways}   Alice\end{sideways}}} & Bits from the RNG & 1     & 0     & 0     & 1     & 0     & 1     & 1     & 1 \bigstrut\\
\cline{2-10}          & Random encoding basis & H-V   & D-A   & H-V   & D-A   & D-A   & H-V   & H-V   & D-A \bigstrut\\
\cline{2-10}          & Photons sent &  $\updownarrow$
   & \reflectbox{\begin{turn}{135}$\leftrightarrow$
\end{turn}}    & $\leftrightarrow$    & \begin{turn}{135}$\leftrightarrow$
\end{turn}    & \reflectbox{\begin{turn}{135}$\leftrightarrow$
\end{turn}}    & $\updownarrow$     & $\updownarrow$    & \begin{turn}{135}$\leftrightarrow$
\end{turn} \bigstrut\\
    \hline
    \multicolumn{1}{|c|}{\multirow{2}[4]{*}{\begin{sideways} Bob\end{sideways}}} & Random decoding basis & H-V   & D-A   & D-A   & D-A   & H-V   & D-A   & D-A   & D-A \bigstrut\\
\cline{2-10}          & Received bits (raw key) & 1     & 0     & 1     & 1     & 0     & 1     & 0     & 1 \bigstrut\\
    \hline
          & Do the bases coincide? & Yes   & Yes   & No    & Yes   & No    & No    & No    & Yes \bigstrut\\
\cline{2-10}          & Sifted key & 1     & 0     &       & 1     &       &       &       & 1 \bigstrut\\
    \hline
    \end{tabular}%
\label{tab1}
\end{table}%
Suppose, then, Eve is performing a simple type of attack, usually referred to as an intercept-resend attack. In this attack, Eve intercepts the train of photons coming from Alice and then sends the obtained qubit sequence to Bob, mimicking Alice. An important point here is the following: since Eve cannot copy the quantum states of original photons, her train of photons will be necessarily different from the one of Alice. Similar to Bob, Eve should also have a passive basis-choice scheme in her measurement instrument. In 50 \% of the cases, she chooses the basis correctly, and then her presence remains unrevealed. However, in the other 50\% of the cases, the photon that Eve sends is polarized on the wrong basis, and thus Bob may obtain the incorrect value of the bit, even if he chooses the right basis according to the public discussion with Alice. Therefore, the active eavesdropping must significantly increase the sifted key error rate, and the presence of Eve could be disclosed if the parties perform an error test. This can be done by comparing some random subsets of the received bits in a public discussion. If the sifted key error rate does not exceed a certain bound, which also takes into account channel noise and setup imperfections, the amount of information available to Eve can be evaluated as not dangerous for the security.  Therefore, the transmission can be considered free of significant eavesdropping, and the shared key could be used for further communication. However, if the amount of errors exceeds the bound, the transmission is considered failed, and the key is disregarded \cite{Bennett2014,Gisin2002}.

If the transmission is considered successful, before the encoding, Alice and Bob perform ordinary procedures of error correction and privacy amplification on the sifted key. Error correction is done to eliminate errors in the sifted key caused by noise in the channel or by the presence of an Eve that could be spoiling a small part of photons without revealing herself. Privacy amplification algorithms are aimed to reduce the information introduced by Eve in the final key \cite{Gisin2002}. 

Note that, like any other QKD protocol, BB84 becomes completely insecure once Eve discovers that the generated random numbers are not truly random and gets the ability to calculate or predict the `random' bits of Alice. Thus, it is essential for generated random numbers in QKD to be truly random. There are plenty of ways to do this, in particular, through quantum random number generators (RNGs) based on an amplified quantum vacuum or an intrinsic probabilistic nature of measurements in the quantum world that seems to be the most trustful source, since the randomness has a scientific proof \cite{Jofre2011, Stipcevic2011}.

Many of the existing QKD systems are based on different modifications of the BB84 protocol. However, the presented scheme is the simplest one, and modern QKD setups are way more complicated. These could be, for instance, the setups that exploit the transverse spatial degree of freedom of photons, which are often referred to as `high-dimensional QKD' \cite{Sit2017, Vallone2014} or the ones based on the phenomenon of quantum entanglement \cite{Ekert1991, Singh2014}.

In an ideal world, communication channels are noiseless. Efficiencies of single-photon sources, photon detectors, and optical elements are unit. Proving the security of a QKD system, in this case, is a straightforward operation. However, real-life physical devices are always imperfect, which could be exploited by Eve to perform attacks on the QKD systems. That is why QKD systems should be designed resistant to different types of eavesdropping attacks.

Single-photon sources, single-photon avalanche photodiodes (SPADs), and other different optical devices (such as beamsplitters, Faraday mirrors, etc.) are the typically used components to build a QKD setup. Each element of the QKD setup has a non-unit efficiency and some inner imperfections, which may open various trapdoors for eavesdroppers. For example, in practice, commercial QKD systems use weak coherent sources instead of single-photon sources \cite{Huang2018}. The latter ones are still under-development, despite the recent progress in single-photon electronics \cite{Aharonovich2016}. A desirable single-photon source at the side of Alice should produce non-classical light, i.e., should emit a pulse of a unique and single-photon once being triggered. Weak coherent sources that are currently available suffer from the probability of emitting several identical photons instead of a single one. An Eve could intercept these `extra' qubits and help to decrypt the information transmitted through the channel without being revealed.

Another problem is that real-life information transfer channels are noisy. As we have previously discussed, the parties could reveal the presence of Eve when the transmission error rate increases in the channel. However, transmission errors may occur in channels that are subject to noise even if there is no eavesdropper. Thus, the question arises of what if Eve that possesses a better technology replaces a part of the channel with a less noisy one? The error rate caused by Eve and the noise in the new channel may be less or equal than the error rate in the primary channel. The presence of Eve could be then disguised as noise.

There is also a wide number of possible quantum attacks based on possible imperfections, as well as a vast number of possible solutions. This gave rise to the relatively new field of research known as quantum hacking, which aims to theoretically and experimentally investigate various types of quantum attacks targeting components of QKD systems that provide Eve with loopholes to secure information.

\begin{table}[htbp]
  \centering
  \caption{Possible Attacks on a QKD System.}
    \begin{tabular}{|l|l|}
    \hline
    Attack  & Description \bigstrut\\
    \hline
    Photon number splitting \cite{Brassard2000} & Current single-photon sources suffer from the ability to\\& emit multiple photons instead of one. The extra photons\\& carry the information and could be intercepted by Eve, causing\\& no errors in the channel. \bigstrut\\
    \hline
    Deadtime \cite{Weier2011} & Each single photon detector has a dead time. By inserting\\& blinding pulses at specific moments, Eve can obtain\\& almost the whole key without being revealed. \bigstrut\\
    \hline
    Trojan-horse \cite{Jain2014} & Eve sends a bright light in the QKD system from the quantum\\& channel and analyzes back reflections. Eve could discern the secret\\& basis choice of Bob and obtain a major part of the key. \bigstrut\\
    \hline
    Laser damage \cite{Bugge2014} & Eve with a high-power laser could damage the QKD system\\& components and alter their characteristics. After that the\\& security of the system may become vulnerable. \bigstrut\\
    \hline
    \end{tabular}%
 \label{tab2}%
\end{table}%

QKD systems are already available on the market today, making quantum cryptography a competitive and fast-growing industry rather than science fiction\cite{comqkd}. However, this rapid growth stirs the interest up for the quantum hacking of such systems. Several practical implementations of different types of successful attacks on commercial QKD systems have been reported during the last few years. Many possible attacks on QKD systems are presented in Table~\ref{tab2}. Further attacks are investigated in \cite{Lydersen2010,Wiechers2011,Lydersen2010a}. These attacks allow one to crack the systems totally, confirming that many existing QKD systems may not be so secure and invulnerable to attacks, opposing to what is stated by several manufacturers. This fact reflects a serious problem in the current market of QKD systems, which is the absence of a single international QKD certification standard. This is mainly because testing a full cryptosystem is very challenging, and therefore it is hard to verify the security levels claimed by manufacturers. Nevertheless, the work on international certification standards is now in progress \cite{etsi}. 

\section{Backflash Problem}
Most of the commercially available QKD setups, today, rely on SPADs as receivers at the side of Bob. SPADs are semiconductor devices that allow Bob to register light radiation at the single-photon level with high efficiency. For telecom wavelengths beyond 1 $\mu$m indium gallium arsenide/indium phosphide (InGaAs/InP) SPADs are used. Silicon-based SPADs (Si-based SPADs) with a larger bandgap are suitable for shorter operating wavelengths, including the visible region of the spectrum. A comparison between several commercially available SPADs in the visible/near-infrared region is presented in Table~\ref{tab3}. The used figures of merit are the peak photon detection efficiency  (PDE$_{\textrm{P}}$) of the SPAD and the dark counting rate (DCR) of the detector. The full-width at half maximum (FWHM) of the distribution diagram characterizes the photon-timing precision of the  SPAD, and $\Phi_{\textrm{M}}$ denotes the maximum achievable photon flux \cite{Bronzi2016}.
\begin{table}[H]
  \centering
  \caption{Several Commercial SPADs Operating in the Visible/Near-Infrared Spectrum Compared.}
    \begin{tabular}{|l|l|l|l|l|c|}
    \hline
    Manufacturer & Model & PDE$_{\textrm{P}}$ [\%] & DCR [cps]   & FWHM [ps] & \multicolumn{1}{l|}{$\Phi_{\textrm{M}}$ [Mcps]} \bigstrut\\
    \hline
    Excelitas Technologies & SPCM-AQRH & 70    & 500   & 350   & \multicolumn{1}{l|}{35} \bigstrut\\
    \hline
    Micro Photon  Devices & PDM   & 50    & 2500  & 35    & \multicolumn{1}{l|}{13} \bigstrut\\
    \hline
    Micro Photon  Devices   & PDM-R & 60    & 2500  & 100   & \multicolumn{1}{l|}{13} \bigstrut\\
    \hline
    ID Quantique & ID 120 & 62    & 150000 & 400  & \multicolumn{1}{l|}{1} \bigstrut\\
    \hline
    Laser Components & COUNT series & 73    & 250   & 800   & \multicolumn{1}{l|}{12} \bigstrut\\
    \hline
    \end{tabular}%
\label{tab3}%
\end{table}%

\begin{figure}[H]
\center{\includegraphics[scale=0.6]{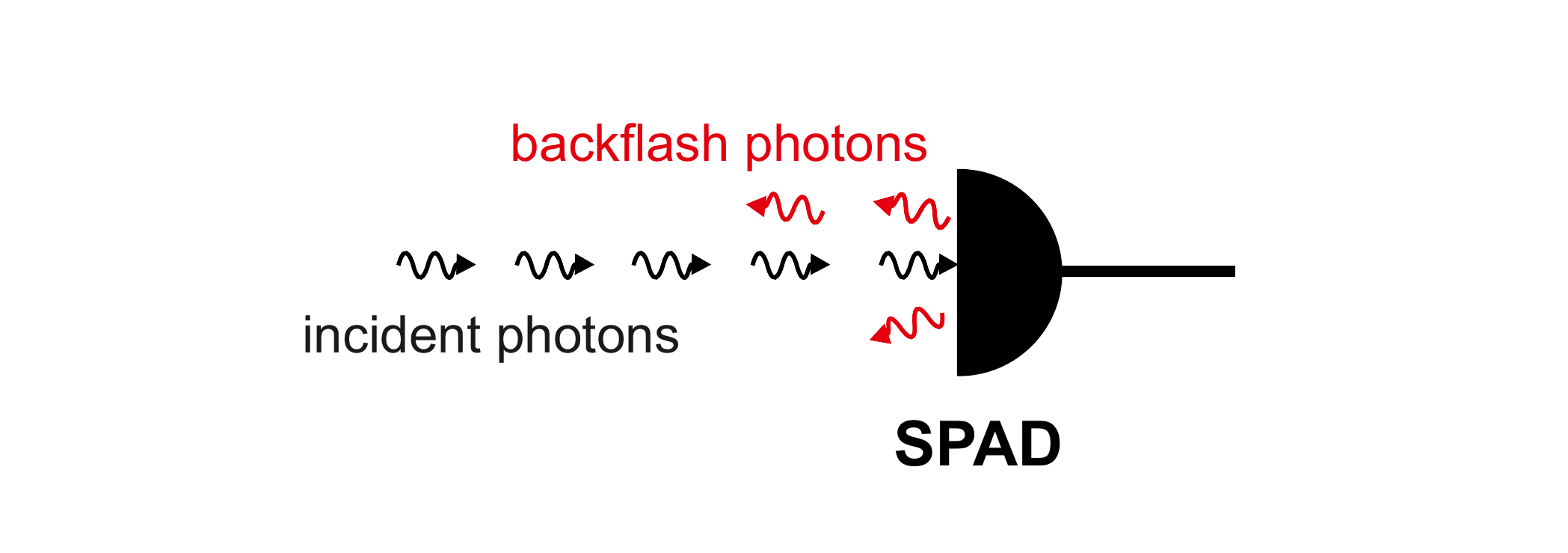}}
\caption{Schematic illustrating the backflash effect.}
\label{Backflash}
\end{figure}

While SPADs have been found to open many trapdoors for eavesdropping attacks \cite{Vakhitov2001,Sajeed2015,Lydersen2010,Gerhardt2011,Li2011}, one  major vulnerability that most SPADs suffer from has been unnoticed for many years and despite the fact that the whole effect was described in the last century, the security threat for QKD was only revealed by recent reports. This vulnerability is known as the backflash light. In fact, all commercially diffused SPADs share the same working principle. SPADs are designed to operate in the so-called Geiger mode, in which the voltage applied to the p-n junction is reversed and goes well beyond the breakdown voltage of the photodiode \cite{Kurtsiefer2001}. In such a situation, a single absorbed photon may trigger a self-sustaining discharge in the SPAD. The avalanche current can then be registered, and the arrival time of the detected photon can be obtained with high timing accuracy. The avalanche current has to be quenched to reset the SPAD and prepare it for the next detection (producing the dead time of the detector) \cite{Hadfield2009}. The quenching is done by a quenching circuit of a certain type, which also affects the SPAD's specifications \cite{Cova1996}. In the 1950s, Newman found that the avalanche of charge carriers during the photon absorption in silicon is accompanied by a significant photon emission \cite{Newman1955}. Early reports have shown that it is the radiative recombination of electrons and holes in the junction that causes this fluorescence light \cite{Gautam1988,Chynoweth1956,Lacaita1993} and gave a quantitative description. This phenomenon is referred to as ``backflash", ``backflash light" or ``breakdown flash". Backflash may provide a side channel to an eavesdropper in certain QKD setups to gain information on photodetection events. A schematic illustrating the emission of backflash photons by a SPAD is depicted in Fig.~\ref{Backflash}.

\begin{figure}[H]
\center{\includegraphics[scale=0.6]{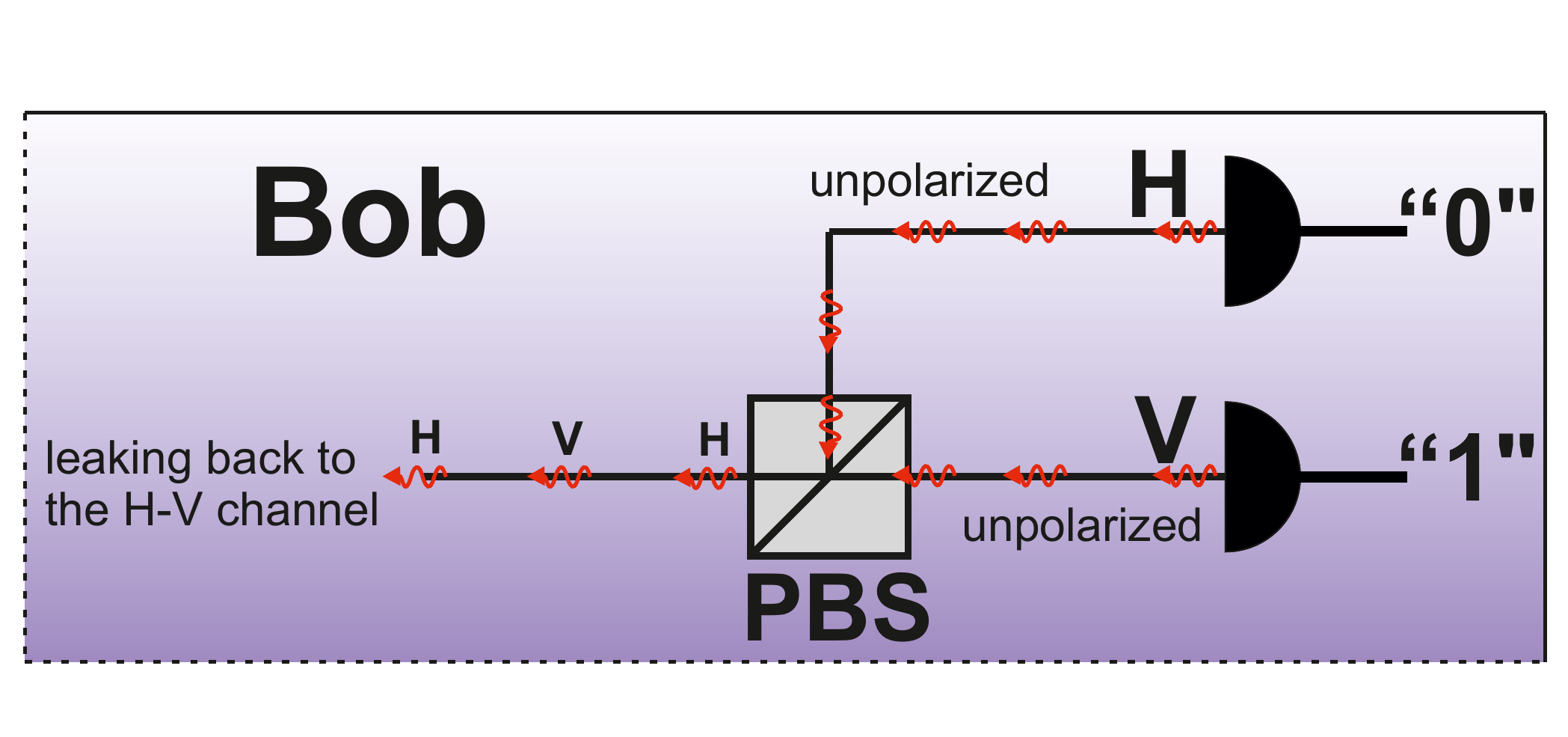}}
\caption{In the BB84 protocol the backflash photons could carry the polarization-encoded information back into the channel.}
\label{BB84P}
\end{figure}

The quantum states of backflash photons seem to be uncorrelated to the state of the absorbed photon. However, since backflash photons could go back through the same polarization-sensitive components of the receiver as the initially detected photons, they can carry the information on photons registered by Bob back to the channel. Such a situation is depicted in Fig.~\ref{BB84P}. Unpolarized backflash photons emerging from the SPADs corresponding to the H and V channels travel back to the PBS. Some of them have a probability of passing through the PBS and getting a polarization following the channel they flew out of. Once an eavesdropper intercepts these photons, the polarization could be decoded, and secret bits may be revealed to the eavesdropper.

In the BB84 QKD scheme, where SPADs and PBSs implement the so-called passive basis-choice scheme, Eve could intercept backflash photons, and measure their polarization state to find out the SPAD (H, V, D or A channel) they flew out of \cite{Pinheiro2018}. The presence of a significant backflash radiation has been experimentally demonstrated in both commercially available InGaAs/InP \cite{Acerbi2013} and Si-based \cite{Kurtsiefer2001} SPADs. In the case of an InGaAs-based SPAD with a nominal detection efficiency of about 10\%, the backflash emission can be a source of significant information leakage, thus compromising the security of the entire QKD system \cite{Shi2017,Meda2016}. An experiment with a silicon-based detector also revealed a considerable rate of backflash emission. However, the estimated information leakage was not high but could be more significant for an Eve with better equipment \cite{Pinheiro2018}.

To access backflash photons, Eve could simply use an optical circulator or a free space optical (FSO) telescope depending on the implementation of the QKD setup \cite{Meda2016,Kupferman2018}. A trapdoor could be open for the so-called zero-error attacks, which refers to the attacks that do not produce errors in the key and therefore are hard to detect. All practical implementations of QKD require the possible information leakage caused by backflash to be accurately quantified. The precise analysis of transmission probabilities and the exact estimation of information leakage bounds for non-ideal media is becoming a hot research topic\cite{Meda2016,Kupferman2018,Zhao2019}.

The properties of backflash light strongly depend on the concrete engineering implementation of the SPAD, particularly, the parameter settings of the quenching electronics \cite{Meda2016}, and on the used semiconductor as well. This explains why different SPAD models operating in different regimes exhibit different backflash spectral distribution, temporal profile, and intensity. Although there exists an analytical theory of backflash radiation, this effect has been mainly investigated experimentally for QKD applications, and the reported results are tied to the properties of the concrete SPADs used \cite{Kurtsiefer2001,Shi2017,Meda2016,Pinheiro2018}.

\begin{figure}[H]
%\label{exp_description}
\center{\includegraphics[scale=0.6]{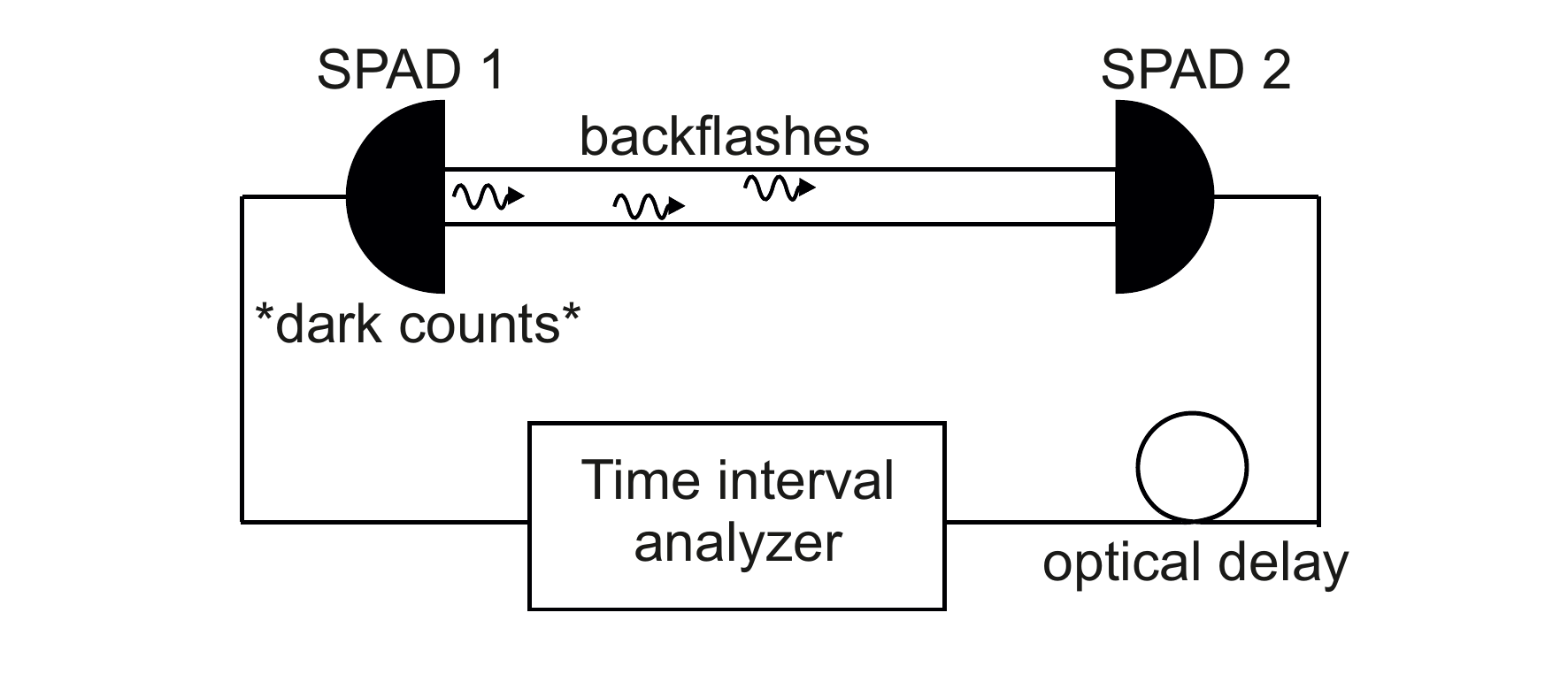}}
\caption{A typical setup for studying the backflash emission properties.}
\label{setup}
\end{figure}

A commonly used setup to quantify backflash radiation is shown in Fig.~\ref{setup}. A pair of SPADs are coupled via an optical fiber or through a line-of-sight (LoS) FSO link. One of the two detectors is chosen to be the under-test device. The backflash radiation that emerged from the chosen SPAD due to the dark counts is studied. The second SPAD aims to register these backflash photons. The coincidences of the counts of both SPADs are inspected using a time interval analyzer to which the outputs of the two detectors are connected. Usually, an optical delay of several dozens of nanoseconds is also added at the output of one of the two SPADs. Measurements performed with such a setup allow one to estimate the probability of backflash, which is the probability that a detection of a photon in the SPAD under test leads to the emission of at least one backflash photon back to the channel. Of course, the non-unit detection efficiency of the second SPAD, as well as optical losses in the channel, should be taken into account during the estimation. The spectral distribution of the backflash light with this setup could be analyzed by plugging different narrow-band filters between the two SPADs or by using diffraction gratings \cite{Shi2017,Pinheiro2018}.

Recent reports have shown that commercial SPADs have unique temporal profiles depending on the detector material and the manufacturer. This fact can potentially allow Eve to identify the type of detector by analyzing the temporal profile of the backflash radiation and then prepare the attacks targeting a particular detector \cite{Meda2016}. Furthermore, for an InGaAs SPAD, authors of \cite{Acerbi2013} experimentally demonstrated that the waveform of avalanche current and the waveform of backflash overlap very well, proving a linear relationship. This has been suggested as a method to estimate the avalanche current waveform of a SPAD non-invasively  \cite{Acerbi2013}. In terms of QKD, variations of such a technique could be probably used by Eve to get additional information on the electronics used in the receiver of Bob.

\begin{figure}[H]
\center{\includegraphics[scale=1.3]{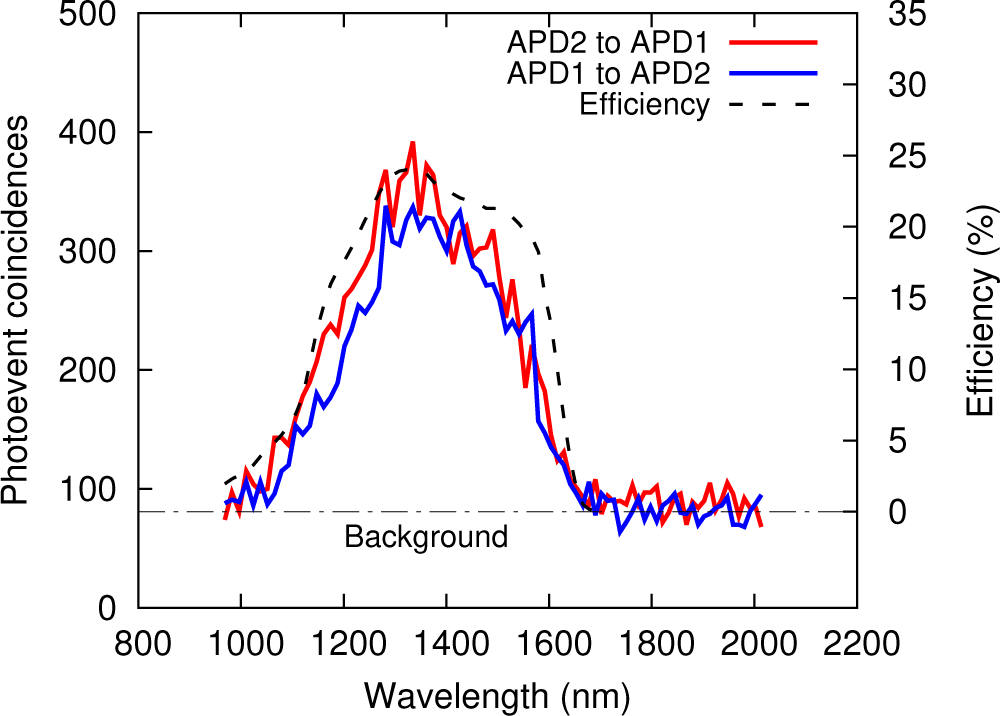}}
\caption{Backflash spectrum of an InGaAs photodiode obtained in \cite{Shi2017}.}
\label{InGaAs_SPAD}
\end{figure}

\begin{figure}[H]
\center{\includegraphics[scale=2.5]{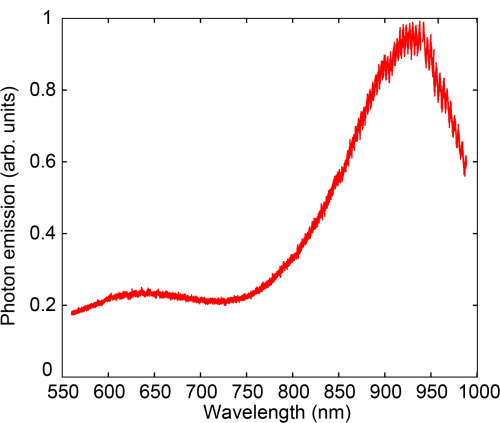}}
\caption{Backflash spectrum of a silicon-based photodiode obtained in \cite{Pinheiro2018}.}
\label{Si_SPAD}
\end{figure}

Backflash photons observed in experiments, however, have relatively broad spectra. The backflash spectrum of an InGaAs SPAD measured in \cite{Shi2017} using a setup similar to the previously described one (see Fig.~\ref{setup}) is presented in Fig.~\ref{InGaAs_SPAD}. The presented results take the background caused by accidental coincidences into account and involve two cases where the first or the second photodiode acts as the under-test device. As can be seen from Fig.~\ref{InGaAs_SPAD}, the spectra cover about 600 nm with a clear peak around 1300 nm. Such a broadband spectrum allows us to suppress significantly, by dozens of times, the rate of backflash by a bandpass filter. A spectrum of backflash radiation emitted by a silicon-based detector, obtained in \cite{Pinheiro2018}, is depicted in Fig.~\ref{Si_SPAD}. Despite that a part of the spectral distribution might cover an area beyond the measurement range, we can see that backflash radiation is broadband and continuous. This means that narrow-pass filtering can be an effective countermeasure against the backflash effect in both cases of SPADs, InGaAs-based, and Si-based.


Another possible countermeasure to the backflash threat is to use optical circulators or optical isolators operating with single photons. These devices at the input of the QKD system could lead backflash photons off the channel to prevent them from being intercepted by Eve. However, we should always keep in mind the non-unit efficiency of such components \cite{Meda2016}. An ultimate solution could be to proceed with photon detection via superconducting-nanowire-based single-photon detectors, which are expected not to produce backflash photons at all. These devices are also available on the market \cite{scdet} and have incomparably higher detection efficiency, lower dark count rate, and shorter dead-time compared to state-of-the-art SPADs. However, the required cryogenics make the cost of such devices too high for being used in commercial QKD systems \cite{Meda2016}.

\section{Future of QKD}
An unprecedented level of security provides QKD with many application opportunities. Although today there are no large-scale quantum computers and existing cryptosystems that can be considered reliable, QKD systems nowadays may be of interest to governments, military agencies, and corporations, which always tend to increase information security levels. Various government and military agencies often act as the main sponsors of QKD research projects, thus confirming their interest in the topic. 

One way to increase the level of data security worldwide using QKD has been recently proposed in \cite{Arnon2019}. The idea is to divide the existing data centers into sub-data centers, which can be connected via optical wireless communication links encrypted using QKD. This will make the penetration and intrusion difficult for hackers, and help to halt the propagation of malware through an entire data center and protect sensitive information.

QKD research worldwide presented a plethora of inspiring demonstrations and experimental results so far, taking the example of the first intercontinental QKD-protected video call demonstration between Beijing (China) and Vienna (Austria) implemented by a low Earth orbit (LEO) satellite connected to two ground stations via LoS optical links \cite{satc,satc2}.  Satellite QKD-based communication is becoming a popular research topic due to the growing interest in LEO satellite constellation systems such as SpaceX Starlink, and OneWeb. These satellite constellations may provide broadband Internet access worldwide, including remote rural areas. Of course, such backhauls will require a high level of cybersecurity, which could be supplied by QKD systems. QKD satellite connections can also be applied for remote surgeries, self-driving cars, and other ambitious urban projects for which security is a crucial aspect. Another significant achievement to mention is the establishment of a quantum backbone network in China connecting Beijing, Jinan, Hefei, and Shanghai, which is a 2000 km multi-node QKD fiber-based network, consisting of 32 trustable relay nodes and 31 fiber links. The customers of this network are the government of China, banks, and some news agencies \cite{mak}. Quantum networks, as well as QKD, are becoming of interest not only for scientists but also for big businesses and governments. QKD networks have also been established in the USA \cite{Hughes2013}, Austria \cite{POPPE2008}, Japan \cite{Sasaki2011}, and Switzerland \cite{Stucki2011} demonstrating high dependability and robustness in real-life environments beyond laboratory test-benches. One promising way to fastly deploy an agile and reconfigurable FSO QKD networks in an urban environment or emergency is to use unmanned aerial vehicles (UAVs). The possible use of modern and stable drones as QKD nodes is being currently investigated \cite{drone}.

Another major challenge of QKD is the relatively high-cost of commercially available equipment. While cost may not be the biggest issue for government and security agencies, it may slow down the global QKD integration in industry and telecommunications systems. However, rapid technological progress may reduce the cost of the current bulky QKD equipment and make it more compact. Eventually, at the beginning of the computer era, 1 Mb of disk storage memory cost about 9000\$, and now it is only about $0.00002$\$ \cite{mem}. So, one day, QKD systems will also become compact and affordable devices contributing to our daily lives.

\bibliography{fileb}
\end{document}